\documentclass[
reprint,
superscriptaddress,
amsmath,amssymb,
aps,
prl,
]{revtex4-2}

\usepackage[version=4]{mhchem}
\usepackage{siunitx}
\usepackage{graphicx}% Include figure files
\usepackage{dcolumn}% Align table columns on decimal point
\usepackage{bm}% bold math
\usepackage{soul}
\usepackage[newcommands]{ragged2e}
\usepackage{caption}
\usepackage{xcolor}
\definecolor{ultramarine}{RGB}{0,112,192}
\usepackage[colorlinks, linkcolor=blue, anchorcolor=blue, citecolor=blue]{hyperref}
\usepackage[utf8]{inputenc}

\begin{document}

%\preprint{APS/123-QED}

\title{Scalable Canonical and Isothermal--Isobaric Sampling of Coupled Spin--Lattice Systems with Machine-Learning Potentials}
%\thanks{A footnote to the article title}

\author{Zhengtao Huang}
\affiliation{Center for Structural Materials, Department of Mechanical Engineering,\\
The University of Hong Kong, Hong Kong, China}

\author{Yunfei Bai}
\affiliation{Graduate School of China Academy of Engineering Physics, Beijing 100088, China}

\author{Han Wang}
\email{wang\_han@iapcm.ac.cn}
\affiliation{HEDPS, CAPT, School of Physics and College of Engineering, \\ 
Peking University, Beijing, 100871, China}
\affiliation{Laboratory of Computational Physics, Institute of Applied Physics and Computational Mathematics, Fenghao East Road 2, Beijing 100094, China}

\author{Ben Xu}
\email{bxu@gscaep.ac.cn}
\affiliation{Graduate School of China Academy of Engineering Physics, Beijing 100088, China}

\date{\today}

\begin{abstract}
% Accurate atomic-scale simulations of materials with coupled spin–lattice degrees of freedom remain hindered by poor energy conservation and prohibitive computational cost, especially with modern machine-learning potentials (MLPs).
% In this work, we introduce TSPIN, a unified Nosé–Hoover chain–based symplectic method that integrates spin and lattice dynamics in NVE, NVT, and NPT ensembles.
% By extending the Lagrangian with spin kinetic terms, TSPIN preserves symplectic structure, ensures robust energy conservation, and reduces model evaluations to one per integration step.
% Benchmarks on harmonic models confirm ensemble accuracy, while large-scale simulations of fcc Fe with a DeepSPIN MLP reveal superior stability and near-linear scaling over conventional LLG.
% This general framework enables robust multi-degree-of-freedom simulations for a broad class of magnetic and correlated materials.
Magnetic machine-learning potentials (MLPs) now reach near-first-principles accuracy on the spin--lattice potential energy surface, but the dynamics and sampling frameworks that convert this accuracy into quantitative finite-temperature thermodynamics have lagged behind.
Landau--Lifshitz--Gilbert spin--lattice dynamics fixes the local moment magnitude and incurs $\mathcal{O}(N)$ MLP evaluations per integration step, while hybrid molecular-dynamics/Monte-Carlo lacks rigorous isothermal--isobaric sampling and remains expensive.
We introduce TSPIN, which promotes the spin to a canonical pair $(\bm S_i,\bm\pi_i)$ alongside the lattice $(\bm R_i,\bm p_i)$ within a Nos\'e--Hoover-chain / Martyna--Tobias--Klein construction, delivering rigorous canonical and isothermal--isobaric sampling, native access to longitudinal spin fluctuations through an unconstrained spin amplitude, and one MLP evaluation per integration step.
Applied to itinerant Co and localized multiferroic \ce{BiFeO3}, TSPIN matches the MD/MC reference thermodynamics of Co at substantially lower cost and reproduces the Curie and N\'eel temperatures within $\sim 7\%$ and $\sim 2\%$ of experiment, respectively. The same unconstrained-amplitude dynamics resolves contrasting spin-amplitude behavior: pronounced spin-modulus softening in Co, but a nearly temperature-independent high-spin Fe$^{3+}$ moment in \ce{BiFeO3}.
TSPIN thereby promotes magnetic MLPs from accurate energy models to predictive finite-temperature simulation engines.
% The method integrates spin and lattice dynamics simultaneously, ensuring energy conservation and reducing computational cost.
% Benchmarks on harmonic models confirm its accuracy, while Co simulations with the DeepSPIN potential demonstrate superior stability, efficiency, and the ability to capture magnetic phase transition. TSPIN provides a general and efficient framework for large-scale simulations of spin–lattice phenomena and multiple-DOF systems. 
% Benchmarks against analytical harmonic spin-lattice models confirm its accuracy, and application to fcc iron using a DeepSPIN machine learning potential demonstrates superior numerical stability and near-linear computational scaling compared to the conventional LLG method. Thus, TSPIN provides a powerful, broadly applicable framework for efficiently simulating complex spin-lattice phenomena and multi-degree-of-freedom systems at large scales.
\end{abstract}

\maketitle

\textit{Introduction}.---Finite-temperature behavior of magnetic materials is set by the joint thermodynamics of spin and lattice degrees of freedom (DOFs), as exemplified by the zero thermal expansion of Fe--Ni Invar alloys~\cite{invar,invar2}, the anomalous phonon lifetimes of paramagnetic CrN~\cite{mdmc}, the temperature-driven spin-reorientation transition in Co~\cite{Co1,Co2}, and the G-type antiferromagnetic order of room-temperature multiferroic \ce{BiFeO3}~\cite{phase2,phase3,BFOCatalan2009}.
Over the past decade, magnetic machine-learning potentials (MLPs)~\cite{aFe,mMTP,npj_noncoll,Ben2023deep,yongxu2024arxiv,hongjun2023arxiv} have pushed the representation of coupled spin--lattice potential energy surfaces to near-first-principles accuracy at tractable cost~\cite{adv_mlp}.
However, improved energy models alone do not guarantee predictive finite-temperature observables: the dynamical and sampling frameworks used to explore these surfaces have advanced more slowly.
As a result, a gap remains between MLP accuracy on the energy surface and the methodology needed to convert that accuracy into quantitative thermodynamics.
Closing this gap requires an atomistic spin--lattice simulation engine that simultaneously delivers (i)~rigorous Boltzmann sampling of a coupled spin--lattice Hamiltonian in both canonical ($NVT$) and isothermal--isobaric ($NpT$) ensembles, (ii)~access to longitudinal spin fluctuations, and (iii)~one model evaluation per integration step.

Existing spin--lattice simulation schemes used with magnetic MLPs fall into two main classes: Landau--Lifshitz--Gilbert (LLG) based spin--lattice dynamics (SLD)~\cite{antropov1996,LSFs,problem2,sl}, which couples Newtonian lattice motion to LLG precession of spins with separate Langevin baths for the two subsystems; and hybrid molecular-dynamics/Monte-Carlo (MD/MC) schemes~\cite{nonsym}, which alternate deterministic lattice MD with Metropolis sampling of spin orientations and, in the most recent magnetic-MLP implementations, of spin magnitudes.
Neither class satisfies requirements~(i)--(iii) simultaneously.
LLG-SLD fixes the magnitude of each local magnetic moment by construction and therefore cannot sample longitudinal spin fluctuations, i.e.\ fluctuations in moment magnitude, that recent magnetic MLPs explicitly resolve, which violates~(ii); its only published $NpT$ variant uses a Berendsen barostat~\cite{LSFs}, which is known not to sample the correct isothermal--isobaric distribution~\cite{tuckerman2010,framework}, violating~(i) in the $NpT$ ensemble; and per-spin Suzuki--Trotter updates incur $\mathcal{O}(N)$ MLP evaluations per integration step~\cite{problem2}, violating~(iii).
MD/MC samples the unconstrained Boltzmann distribution, including fluctuations in spin magnitude, satisfying~(i) and~(ii) for $NVT$, but operates at fixed volume with thermal expansion handled \emph{post hoc} via stress--strain integration~\cite{nonsym}, thereby violating~(i) for the $NpT$ sampling, and requires a separate equilibration run at each temperature point with combined MD and MC overhead far exceeding one model evaluation per step, violating~(iii).

Here we introduce TSPIN, a unified spin--lattice simulation framework in which the spin vector $\bm{S}_i\in\mathbb R^3$ is treated as an independent atomic degree of freedom with conjugate momentum $\bm{\pi}_i$, and the coupled spin--lattice Hamiltonian is extended into the Nos\'e--Hoover chain (NHC)~\cite{nose} thermostat and Martyna--Tobias--Klein (MTK)~\cite{nose_npt,framework} barostat construction.
This generalises the Nos\'e--Hoover spin thermostat of Antropov \emph{et al.}~\cite{antropov1996}, which operated on the fixed-$|\bm{S}|$ manifold, to a magnitude-unconstrained spin degree of freedom inside the full NHC/MTK construction.
This single architectural choice delivers requirements~(i)--(iii) at once: the measure-preserving NHC/MTK construction~\cite{framework,tuckerman2010} samples the coupled spin--lattice Hamiltonian rigorously in both $NVT$ and $NpT$, satisfying~(i); the unconstrained spin amplitude grants direct access to longitudinal spin fluctuations (LSFs), satisfying~(ii); and the single deterministic integrator requires one MLP evaluation per integration step, satisfying~(iii).
We benchmark TSPIN on a harmonic spin--lattice model, confirming canonical sampling across a wide temperature range.
Applied to itinerant Co, TSPIN reproduces the MD/MC reference thermodynamics~\cite{nonsym} at one MLP evaluation per step and quantitatively captures the ferromagnetic--paramagnetic transition.
For localized multiferroic \ce{BiFeO3}, it predicts the G-type antiferromagnetic N\'eel temperature in agreement with experiment.
TSPIN thereby promotes magnetic MLPs from accurate energy models to predictive finite-temperature simulation engines for magnetic phase transitions at large atomistic scale, with longitudinal spin fluctuations and the isothermal--isobaric ensemble natively supported by the construction.

\textit{Theoretical framework}.---Consider a classical system of $N$ magnetic particles with lattice coordinates $\{\bm{R}_i\}$ and spin vectors $\{\bm{S}_i\}$, $\bm{R}_i, \bm{S}_i \in \mathbb{R}^3$, and potential energy $U(\{\bm{R}\},\{\bm{S}\})$.
The atomic forces and magnetic torques follow as $\bm{F}_i = -\partial U/\partial \bm{R}_i$ and $\bm{\omega}_i = -\partial U/\partial \bm{S}_i$, respectively.
To treat the spin DOFs on equal footing with the lattice coordinates and cast the coupled spin--lattice problem in canonical Hamiltonian form, we introduce a fictitious spin inertia $\mu_i$~\cite{CPMD,spin_inertia} and conjugate momentum $\bm{\pi}_i$, elevating $(\bm{S}_i, \bm{\pi}_i)$ to a canonical pair alongside the lattice pair $(\bm{R}_i, \bm{p}_i)$.

The microcanonical ($NVE$) Hamiltonian is
\begin{equation}
\mathcal{H}_0 = \sum_i \frac{\bm{p}_i^2}{2M_i} + \sum_i \frac{\bm{\pi}_i^2}{2\mu_i} + U(\{\bm{R}\},\{\bm{S}\}),
\label{eq:h0}
\end{equation}
with equations of motion $\dot{\bm{R}}_i = \bm{p}_i/M_i$, $\dot{\bm{p}}_i = \bm{F}_i$, $\dot{\bm{S}}_i = \bm{\pi}_i/\mu_i$, and $\dot{\bm{\pi}}_i = \bm{\omega}_i$.
For $U$ invariant under spatial translation and joint rotation of lattice and spin variables, energy, linear momentum, and total angular momentum $\bm{J} = \sum_i (\bm{R}_i\times\bm{p}_i + \bm{S}_i\times\bm{\pi}_i)$ are conserved (Supplemental Material~\cite{SM}, Sec.~I).

To sample the canonical distribution, the lattice and spin DOFs are coupled to a Nos\'e--Hoover chain (NHC) thermostat~\cite{nose}, yielding the extended Hamiltonian
\begin{equation}
\mathcal{H}_{NVT} = \mathcal{H}_0 +
\sum_{k=1}^{C}\frac{p_{\xi_k}^2}{2Q_k} + Lk_BT\,\xi_1 + \sum_{k=2}^{C}k_BT\,\xi_k,
\label{eq:hnvt}
\end{equation}
where $\xi_k$, $p_{\xi_k}$, and $Q_k$ are the coordinates, momenta, and masses of the NHC of length $C$, and $L = 2dN$ is the total number of physical DOFs.
The lattice and spin momenta evolve as
\begin{equation}
\dot{\bm{p}}_i = \bm{F}_i - \frac{p_{\xi_1}}{Q_1}\bm{p}_i, \qquad \dot{\bm{\pi}}_i = \bm{\omega}_i - \frac{p_{\xi_1}}{Q_1}\bm{\pi}_i.
\label{eq:eom_nvt}
\end{equation}
The remaining NHC equations of motion for the chain variables follow their standard form and are given in Supplemental Material~\cite{SM}, Sec.~II.

The extended-system flow is non-Hamiltonian but measure-preserving, with metric $\sqrt{g}=\exp(L\,\xi_1+\xi_c)$, $\xi_c\equiv\sum_{k=2}^{C}\xi_k$; integrating the chain and momentum variables out of the generalised microcanonical density yields the configurational canonical distribution $\rho(\bm{R},\bm{S}) \propto e^{-\beta U}$~\cite{framework,tuckerman2010} (see Supplemental Material~\cite{SM}, Sec.~II).
The auxiliary inertia $\mu_i$ does not affect this distribution, since integration over $\bm{\pi}_i$ contributes only a constant.
It serves as a Car--Parrinello-style timescale parameter~\cite{CPMD}, chosen to render spin and lattice timescales comparable.
It may also be interpreted as an effective spin--bath inertia~\cite{spin_inertia}.

To sample the isothermal--isobaric distribution, the $NVT$ Hamiltonian is extended by a fluctuating volume $V$ with conjugate momentum $p_\epsilon$, barostat mass $W$, and external pressure $P_{\rm ext}$, giving the $NpT$ Hamiltonian
\begin{equation}
\mathcal{H}_{NpT} = \mathcal{H}_{NVT}\big|_{L \to L+1} + \frac{p_\epsilon^2}{2W} + P_{\rm ext}V.
\label{eq:hnpt}
\end{equation}
The Martyna--Tobias--Klein (MTK) construction~\cite{nose_npt} adds barostat coupling to the lattice DOFs; the spin equations are unchanged from $NVT$ (Supplemental Material~\cite{SM}, Sec.~III).
The same generalised-Liouville analysis extends to the volume DOF, yielding the configurational isothermal--isobaric distribution $\rho(\bm{R},\bm{S},V) \propto e^{-\beta[U(\bm{R},\bm{S}) + P_{\rm ext}V]}$.
MTK is the canonical $NpT$ MD construction proved to yield the correct distribution~\cite{framework,tuckerman2010}.

For both $NVT$ and $NpT$, the equations of motion are integrated by a symmetric Suzuki--Trotter factorisation of the Liouville operator (Supplemental Material~\cite{SM}, Sec.~IV), giving a time-reversible and measure-preserving, second-order integrator~\cite{framework}.

\textit{Benchmark on a harmonic potential}.---We first verify that TSPIN samples the correct canonical distribution on an analytically tractable model: $N=1000$ non-interacting magnetic particles with
$U(\bm{R},\bm{S}) = \sum_{i=1}^N\!\left(\tfrac{1}{2}k|\bm{R}_i|^2 + \tfrac{1}{2}w|\bm{S}_i|^2\right)$,
for which the equilibrium marginals of $\bm{R}_i$, $\bm{p}_i$, $\bm{S}_i$, and $\bm{\pi}_i$ are Gaussians with variances set by equipartition.
$NVT$ simulations are performed in a cubic cell using a modified version of LAMMPS~\cite{lammps},
with atomic mass $M_i=57.5\,\mathrm{g/mol}$, spin inertia $\mu_i=0.01\,\mathrm{eV\cdot ps^2/\mu_B^2}$, NHC length $C=3$, and harmonic constants $k=1.75\,\mathrm{eV/\text{\AA}^2}$, $w=0.75\,\mathrm{eV/\mu_B^2}$.
The equations of motion are integrated with a time step of $1\,\mathrm{fs}$ for $2\times10^5$ steps ($200\,\mathrm{ps}$).
% The simulations are carried out with atomic mass $M_i = 57.5$, spin mass $\mu_i = 0.01$, NHC length $C = 3$, and harmonic constants $k = 1.75$ and $w = 0.75$. The dynamics are integrated with a 0.1 fs time step over 2,000,000 steps (200 ps).
% Equilibrium distributions for lattice positions, momenta, and spins are extracted and compared against analytical results.

\begin{figure}[tt]
\centering
\includegraphics[width=1.0\linewidth]{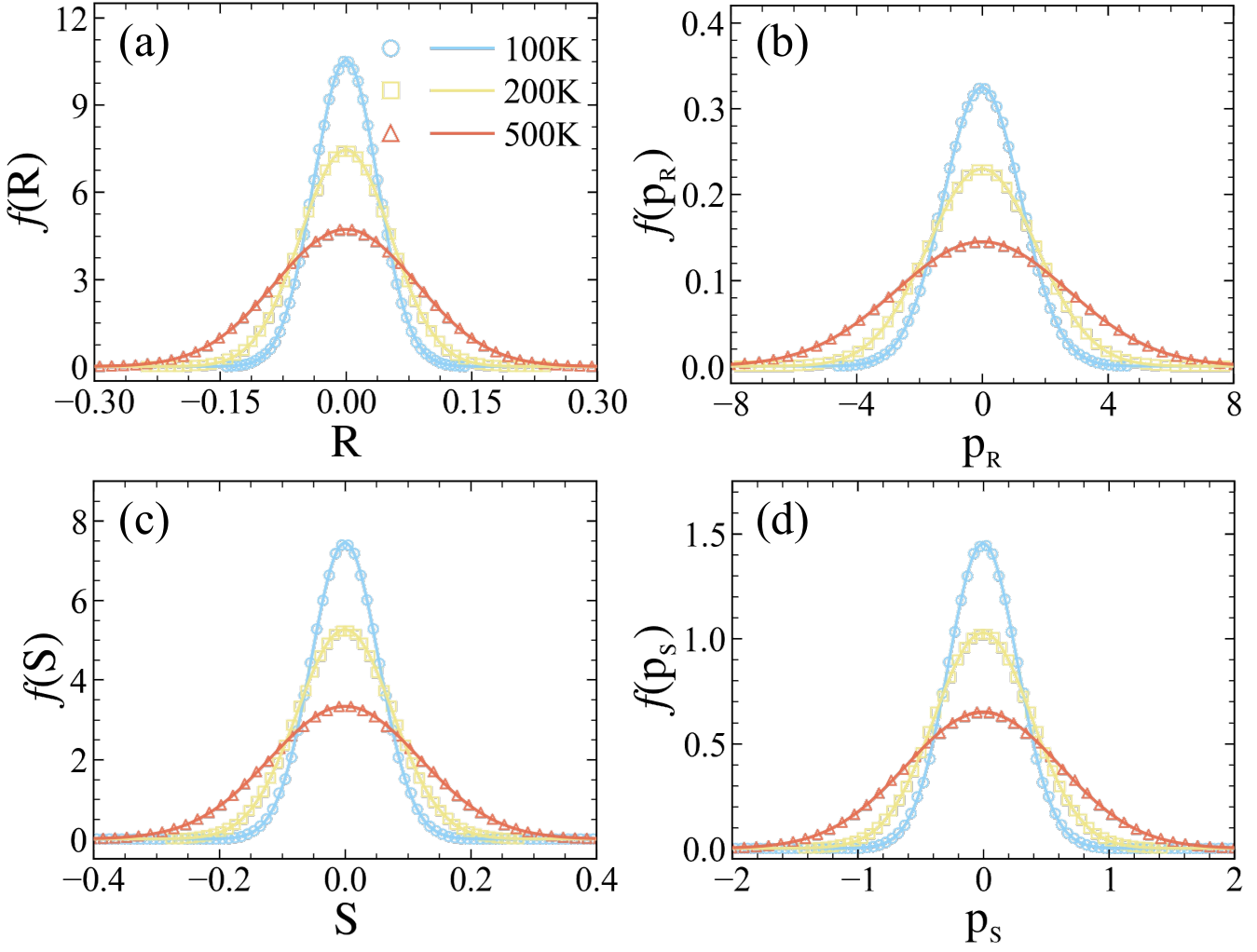}
\caption{Benchmark of TSPIN on a non-interacting harmonic spin--lattice model at $T=100$, $200$, and $500\,\mathrm{K}$. Panels (a)--(d) show the equilibrium distributions of the lattice coordinate $\bm{R}_i$, lattice momentum $\bm{p}_i$, spin $\bm{S}_i$, and spin momentum $\bm{\pi}_i$, respectively. Symbols: simulation; solid lines: analytical Gaussians.}
\label{fig:spin_dis} 
\end{figure}

Figure~\ref{fig:spin_dis} compares the simulated equilibrium distributions of the lattice and spin variables at $T=100$, $200$, and $500\,\mathrm{K}$ against the analytical Gaussians.
The simulation results (symbols) match the analytical distributions (solid lines) across all three temperatures,
confirming that TSPIN samples the correct canonical distribution for the spin--lattice system.

% 思考下补充材料放什么内容? 1. 公式推导. 2. model构建 数据量 RMSE等. temp 数据量 RMSE.
% 图片调整为自上而下的排列.

\textit{Performance in cobalt}.---To demonstrate the applicability of TSPIN to realistic magnetic materials, we consider cobalt (Co), a prototypical itinerant ferromagnet whose strong spin--lattice coupling drives a temperature-driven ferromagnetic--paramagnetic transition~\cite{Co1,Co2,Co3}.
The interactions are described by the DeepSPIN model~\cite{Ben2023deep}, which extends DeepPot-SE~\cite{deepmd} to magnetic systems via a pseudo-atom representation of spins; its Co-specific parameterisation is trained with the concurrent-learning framework DPGEN2~\cite{dpgen} over $0$--$1600\,\mathrm{K}$, with construction details in Supplemental Material~\cite{SM}, Sec.~V.

\begin{figure}[tt]
\centering
\includegraphics[width=1.0\linewidth]{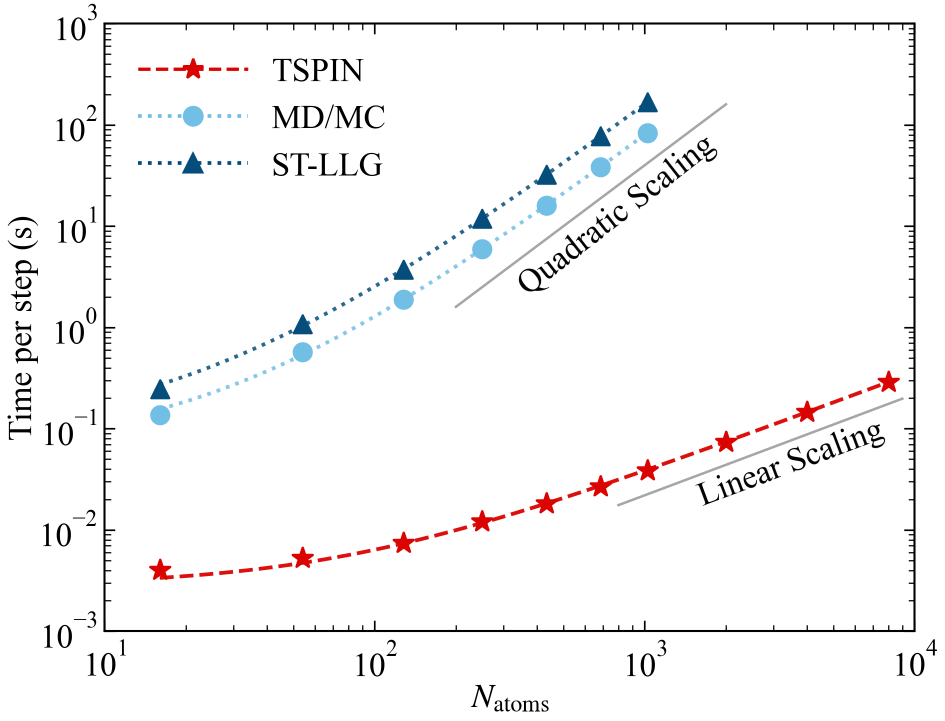}
\caption{Per-step wall-clock time of TSPIN, ST-LLG~\cite{problem2}, and hybrid MD/MC~\cite{mdmc,nonsym} on hcp Co as a function of system size. Measurements on a single NVIDIA V100 GPU using a modified version of LAMMPS.}
\label{fig:time_cost}
\end{figure}

We first quantify the computational cost of TSPIN relative to LLG-based dynamics and hybrid MD/MC, as this sets the accessible system size for direct method-to-method comparisons.
Figure~\ref{fig:time_cost} reports the per-step wall-clock time on an NVIDIA V100 GPU as a function of system size.
TSPIN scales linearly with $N_{\mathrm{atoms}}$, requiring only a single MLP evaluation per integration step, as in classical MD.
For LLG we adopt the symplectic per-spin Suzuki--Trotter LLG integrator (ST-LLG)~\cite{problem1,problem2} as implemented in the LAMMPS spin package and widely used for spin--lattice MD simulations; its sequential per-spin updates invoke the MLP $4N$ times per step~\cite{problem2}, giving quadratic wall-clock scaling.
Hybrid MD/MC likewise scales quadratically, as each Monte Carlo sweep requires $\mathcal{O}(N)$ energy evaluations and convergence demands many alternating iterations.
These scalings limit direct three-method comparisons to modest system sizes, motivating the $4\times4\times4$ supercell used for the benchmark below.

We perform $NVT$ simulations of hexagonal close-packed (hcp) Co on a $4\times4\times4$ supercell (128 atoms), initialised in a ferromagnetic state with spins along the $c$-axis, at $T=200$, $400$, and $600\,\mathrm{K}$, with atomic mass $M=58.93\,\mathrm{g/mol}$ and spin inertia $\mu=0.01\,\mathrm{eV\cdot ps^2/\mu_B^2}$, and compare against hybrid MD/MC~\cite{mdmc,nonsym} (which samples the joint distribution $\rho(\bm R,\bm S)$ by alternating lattice MD with spin MC and serves as the reference) and against ST-LLG.

\begin{figure}[tt]
\centering
\includegraphics[width=1.0\linewidth]{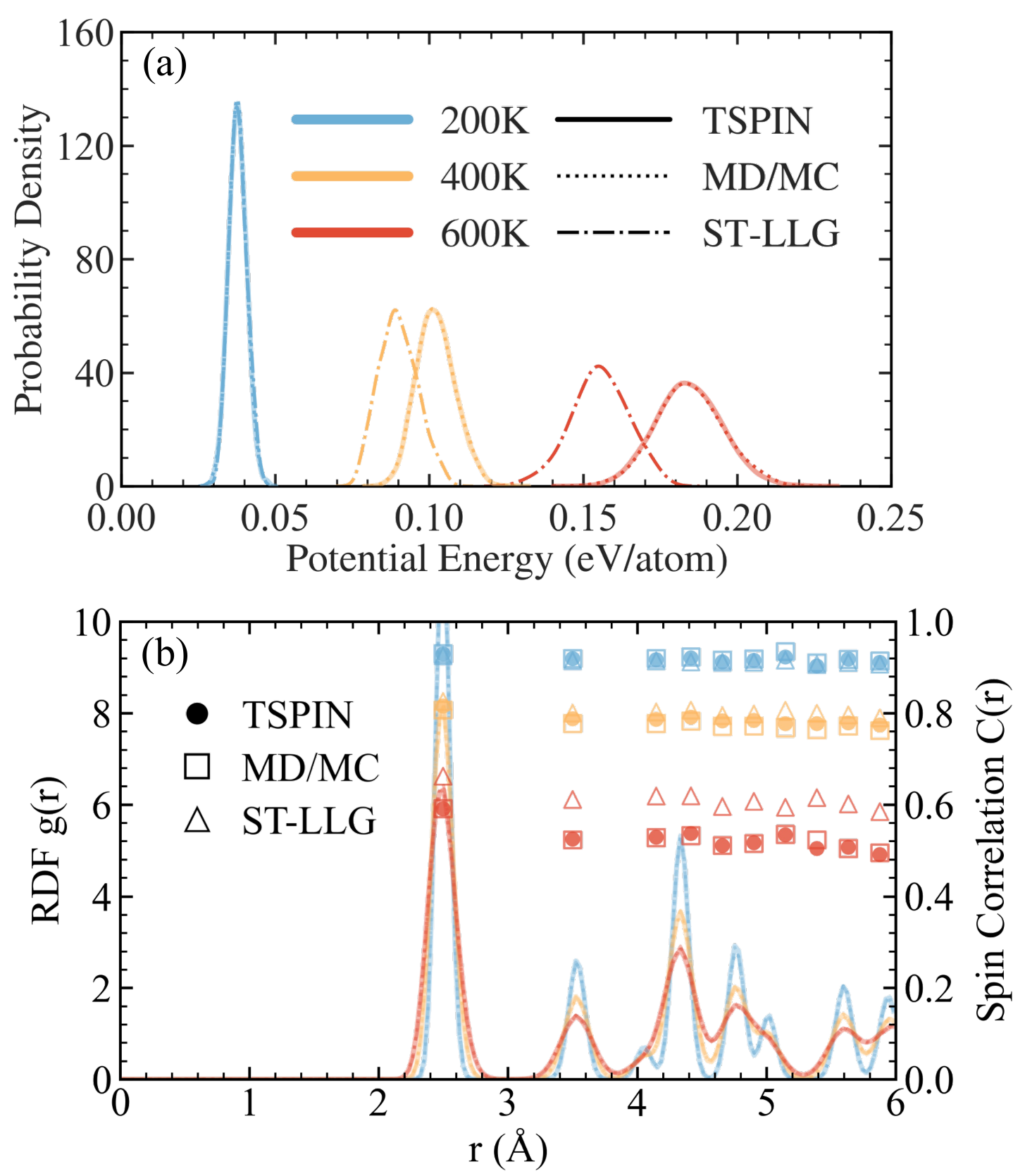}
\caption{Equilibrium properties of hcp Co at $T=200\,\mathrm{K}$ (blue), $400\,\mathrm{K}$ (orange), and $600\,\mathrm{K}$ (red) from TSPIN (solid), MD/MC (dotted), and ST-LLG (dash-dotted). (a) Potential-energy probability density. (b) Lattice radial distribution function $g(r)$ (lines) and radial spin correlation $C(r)$ (symbols) versus interatomic distance $r$.}
\label{fig:drift} 
\end{figure}

As shown in Fig.~\ref{fig:drift}(a), the potential-energy distributions from TSPIN are nearly identical to those from MD/MC at all three temperatures.
The agreement extends to the lattice radial distribution function $g(r)$ and the radial spin correlation $C(r) = \langle\bm{S}_i\cdot\bm{S}_j\rangle_{|\bm r_{ij}|=r}/\langle S^2\rangle$~\cite{spincorr} [Fig.~\ref{fig:drift}(b)], confirming accurate sampling of the coupled spin--lattice equilibrium.
ST-LLG, by contrast, samples a restricted distribution $\rho(\bm R,\bm S)\,\delta(|\bm S|-S_0)$ because the fixed-magnitude constraint neglects longitudinal spin fluctuations (LSFs)~\cite{LSFs,LSFs2}, which TSPIN and MD/MC naturally capture.
At $200\,\mathrm{K}$ LSFs remain weak and ST-LLG reproduces the reference energy and correlation functions; at $400$ and $600\,\mathrm{K}$ the fixed-magnitude constraint progressively underestimates the potential-energy distribution [Fig.~\ref{fig:drift}(a)] and overestimates $C(r)$ even though $g(r)$ remains accurate [Fig.~\ref{fig:drift}(b)], consistent with the growing importance of LSFs at elevated temperatures.

\begin{figure}[t]
\centering
\includegraphics[width=1.0\linewidth]{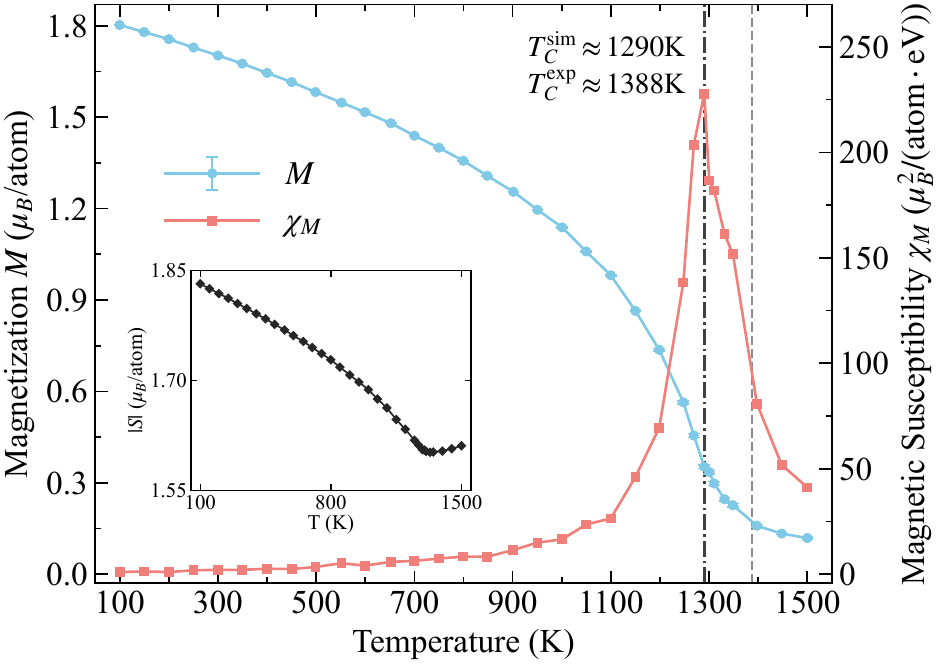}
\caption{
Curie transition in itinerant fcc Co. Temperature dependence of the per-atom magnetization $M$ (blue, left axis) and magnetic susceptibility $\chi_M$ (red, right axis); symbols are simulation data and error bars indicate statistical uncertainty. The dash-dotted line marks the simulated Curie temperature $T_C^{\mathrm{sim}}\approx1290\,\mathrm{K}$ from the $\chi_M$ peak; the dashed line marks the experimental value $T_C^{\mathrm{exp}}\approx1388\,\mathrm{K}$. Inset: average spin modulus $|S| = \frac{1}{N}\sum_{i}\langle|\bm S_i|\rangle$.
}
\label{fig:co_curie}

\vspace{1.0\baselineskip}

\centering
\includegraphics[width=1.0\linewidth]{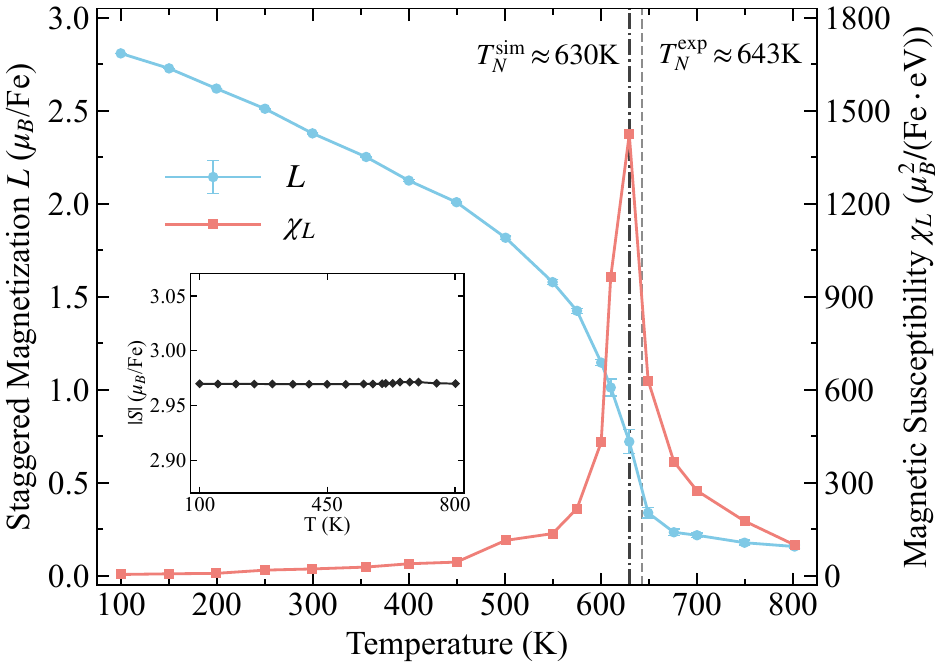}
\caption{
N\'eel transition in multiferroic \ce{BiFeO3}.
Temperature dependence of the staggered magnetization $L$ and susceptibility $\chi_L$.
% The vertical dash-dotted and dashed lines mark $T_N^{\mathrm{sim}}\approx630\,\mathrm{K}$ from the susceptibility peak and the experimental value $T_N^{\mathrm{exp}}\approx643\,\mathrm{K}$~\cite{BFOCatalan2009,BFOXu2012}, respectively.
Inset: average Fe spin modulus $|S|$ over the same temperature range, with $|S| = \frac{1}{N_{\mathrm{Fe}}}\sum_{i\in\mathrm{Fe}}\langle|\bm S_i|\rangle$.
}
\label{fig:bfo_neel}
\end{figure}

The ferromagnetic--paramagnetic (Curie) transition of face-centered-cubic (fcc) Co, with experimental critical temperature $T_C^{\mathrm{exp}}=1388\,\mathrm{K}$~\cite{Kittel2005}, provides a direct benchmark.
We perform $NpT$ simulations on an $8\times8\times8$ supercell ($2{,}048$ atoms) at zero external pressure across $T=100$--$1500\,\mathrm{K}$, computing the magnetization $M=\langle|\bm m|\rangle$ with $\bm m=\tfrac{1}{N}\sum_i\bm S_i$ and the susceptibility $\chi_M=\frac{N}{k_BT}\!\left(\langle|\bm m|^2\rangle-\langle|\bm m|\rangle^2\right)$ from its equilibrium fluctuations.
Figure~\ref{fig:co_curie} shows $M(T)$ decreasing monotonically and dropping sharply near a pronounced $\chi_M$ peak at $T_C^{\mathrm{sim}}\approx 1290\,\mathrm{K}$, within $\sim 7\%$ of $T_C^{\mathrm{exp}}$.
The inset reveals a strong temperature dependence of the average spin modulus $|S|$, contrasted with the \ce{BiFeO3} case below; the full $|\bm S_i|$ distributions are provided in Supplemental Material~\cite{SM}, Sec.~VI.

\textit{N\'eel transition in BiFeO$_3$}.---Rhombohedral \ce{BiFeO3} is a room-temperature multiferroic that undergoes a G-type antiferromagnetic--paramagnetic (N\'eel) transition at $T_N^{\mathrm{exp}}\approx 643\,\mathrm{K}$~\cite{BFOCatalan2009,BFOXu2012}, providing a complementary benchmark.
We perform $NVT$ simulations on a $10\times10\times10$ supercell ($10{,}000$ atoms, $2{,}000$ Fe), initialised in the G-type antiferromagnetic state, across $T=100$--$800\,\mathrm{K}$, with a $100\,\mathrm{ps}$ trajectory per temperature, $1.0\,\mathrm{fs}$ timestep, and spin inertia $\mu=0.01\,\mathrm{eV\cdot ps^2/\mu_B^2}$.
The \ce{BiFeO3} DeepSPIN potential is trained with the same DPGEN2~\cite{dpgen} concurrent-learning protocol as for Co; see Supplemental Material~\cite{SM}, Sec.~V, for construction details.

We characterise the antiferromagnetic order through the Fe-sublattice staggered magnetization $\bm L = \frac{1}{N_{\mathrm{Fe}}}\sum_{i\in\mathrm{Fe}}\eta_i\bm S_i$, where $\eta_i=\pm 1$ labels the two G-type sublattices, with scalar order parameter $L=\langle|\bm L|\rangle$.
The staggered susceptibility is computed from equilibrium fluctuations~\cite{landau_binder2014}, $\chi_L=\frac{N_{\mathrm{Fe}}}{k_BT}\!\left(\langle|\bm L|^2\rangle-\langle|\bm L|\rangle^2\right)$.
Figure~\ref{fig:bfo_neel} shows $L(T)$ decreasing continuously and a sharp $\chi_L$ peak at $T_N^{\mathrm{sim}}\approx 630\,\mathrm{K}$, within $\sim 2\%$ of $T_N^{\mathrm{exp}}$ and stable with respect to supercell size (Supplemental Material~\cite{SM}, Sec.~VII).

Co and \ce{BiFeO3} jointly test TSPIN's unconstrained-amplitude dynamics in two opposing regimes, itinerant and localized.
In itinerant Co, the average spin modulus $|S|$ softens by $\sim15\%$ on approaching $T_C$ and remains finite in the paramagnetic phase [Fig.~\ref{fig:co_curie} inset], indicating LSFs.
In localized \ce{BiFeO3}, by contrast, the Fe spin modulus $|S|$ remains nearly independent of $T$ across $T_N$ [Fig.~\ref{fig:bfo_neel} inset], consistent with high-spin Fe$^{3+}$ ($d^5$) moments; the loss of staggered order is therefore dominated by orientational disordering.
Both regimes are captured by the same TSPIN dynamics, with no per-system retuning: each $\bm S_i$ evolves freely, and the resulting $|S|(T)$ is determined by the spin--lattice free energy rather than by a prescribed constraint or sampling protocol.

\textit{Conclusions}.---
In this Letter, we introduced TSPIN, which promotes the spin to a canonical pair $(\bm S_i,\bm\pi_i)$ alongside the lattice $(\bm R_i,\bm p_i)$ and develops the coupled equations of motion for both spin and lattice DOFs in the $NVT$ and $NpT$ ensembles.
This single architectural choice resolves the three obstacles that prevent magnetic MLPs from yielding quantitative finite-temperature thermodynamics: rigorous sampling of Boltzmann distributions, native access to longitudinal spin fluctuations, and one MLP evaluation per integration step.
TSPIN matches MD/MC equilibrium thermodynamics at substantially lower cost and quantitatively reproduces both the experimental Curie temperature of Co ($T_C^{\mathrm{sim}}\approx 1290\,\mathrm{K}$, within $\sim 7\%$ of $T_C^{\mathrm{exp}}$) and the N\'eel temperature of multiferroic \ce{BiFeO3} ($T_N^{\mathrm{sim}}\approx 630\,\mathrm{K}$, within $\sim 2\%$ of $T_N^{\mathrm{exp}}$).
The same dynamics captures both itinerant and localized regimes with no per-system retuning, $|\bm S_i|(T)$ emerging from the spin--lattice free energy.
TSPIN thus turns magnetic MLPs from accurate energy models into predictive finite-temperature simulation engines, opening the way to large-scale studies of magneto-structural coupling, magnetic phase transitions, and other strongly coupled multi-DOF systems.

\textit{Code availability}.---The modified LAMMPS implementation of the TSPIN
spin--lattice integrator is available at \url{https://github.com/hztttt/lammps_NHC_SPIN}.

\begin{acknowledgments}
B.X. acknowledges funding from Science Challenge Project No.TZ2025017. The work of Han Wang is supported by the National Natural Science Foundation of China (Grants No.~12525113 and No.~12561160120).
\end{acknowledgments}

\bibliographystyle{unsrt}
\bibliography{main.bib}

\end{document}